\documentstyle[aps,prd,eqsecnum,preprint]{revtex}
\begin{document}
\title{\hfill OKHEP-96-04\\
Casimir Energy for a Spherical Cavity in a
Dielectric:\\Applications to Sonoluminescence}
\author{Kimball A. Milton\thanks{e-mail: kmilton@ou.edu}}
\address{ Department of Physics and Astronomy,
The University of Oklahoma, Norman OK 73019 USA}
\author{Y. Jack Ng\thanks{e-mail: ng@physics.unc.edu}}
\address{Institute of Field Physics, Department of Physics and
Astronomy,
University of North Carolina, Chapel Hill, NC 27599 USA}
\date{\today}
\maketitle

\begin{abstract}
In the final few years of his life, Julian Schwinger proposed that
the ``dynamical Casimir effect'' might provide the driving force
behind the puzzling phenomenon of sonoluminescence.  Motivated by
that exciting suggestion, we have computed the static Casimir energy
of a spherical cavity in an otherwise uniform material.
 As expected the result is divergent;  yet a plausible finite answer
is
 extracted, in the leading uniform asymptotic approximation.
This result agrees with that found using $\zeta$-function
regularization.
Numerically, we find far too small an
energy to account for the large burst of photons seen in
sonoluminescence.
If the divergent result is retained,
it is of the wrong sign to drive the effect.
Dispersion does not resolve this contradiction.
In the static approximation, the Fresnel drag term is zero; on the
other hand, electrostriction could be comparable to the Casimir term.
It is argued that this adiabatic approximation to the dynamical
Casimir
effect should be quite accurate.
\end{abstract}

\section{Introduction}
In a series of papers in the last three years of his life, Julian
Schwinger proposed \cite{js} that the dynamical Casimir effect could
provide the energy that drives the copious production of photons
in the puzzling phenomenon of sonoluminescence
\cite{sono1,sono2,isotope}.
In fact, however, he guessed an approximate (static) formula for the
Casimir energy of a spherical bubble in water, based on a general,
but incomplete, analysis \cite{rederiv}.  He apparently
was unaware that one of us had, in the late 1970's, completed the
analysis of the Casimir force for a dielectric ball \cite{kim}.
It is our purpose here to carry out the very straightforward
calculation for the complementary situation, for a cavity in an
infinite dielectric medium.\footnote{A
preliminary version of this paper
appeared in \protect{\cite{leipzig}}.}
  In fact, we will consider the general
case of spherical region, of radius $a$, having permittivity
$\epsilon'$ and permeability $\mu'$, surrounded by an infinite
medium of permittivity $\epsilon$ and permeability $\mu$.

Of course, this calculation is not directly relevant to
sonoluminescence, which is anything but static.  It is
offered as only a preliminary step, but it
should give an idea of the orders of magnitude of the energies
involved. It is a significant improvement over the crude
estimation used in \cite{js}. Attempts at dynamical calculations
exist \cite{sass,eberlein,chodos};
but they are subject to possibly serious
methodological objections, some of which will be discussed below.
(Other theoretical models to explain sonoluminescence are given in
\cite{theory}.)
In fact, we anticipate that because the relevant scale of the
electromagnetic
Casimir effect is in the optical region, with characteristic time
scale $t\sim 10^{-15}$s, and the scale of the bubble collapse is
of order $\tau\sim 10^{-6}$s (more relevant may be the duration of
each flash, which is $\lesssim 10^{-11}$s), the adiabatic
approximation of treating the bubble as static for calculating
the Casimir energy should be very accurate.
Sonoluminescence aside, this calculation is of interest for
its own sake, as one of a relatively few nontrivial Casimir
calculations with nonplanar boundaries
\cite{boyer,balian,mds,dm,bn,fermion,mng,bm}.  It represents a
significant generalization on the calculation of Brevik and
Kolbenstvedt \cite{brevik}, who consider the same geometry with
 $\mu\epsilon=\mu'\epsilon'=1$, a special case, possibly relevant
to hadronic physics, in which the result
is unambiguously finite.  It is, as noted above, a straightforward
generalization of the result in \cite{kim}; the most significant
technical
improvement is that here the energy is calculated directly.
We also examine Fresnel drag and electrostriction; the latter may
be numerically significant.

In the next section we review the Green's dyadic formalism we shall
employ, and compute the Green's functions in this case for the TE and
TM modes.  Then, in Section 3, we compute the force on the cavity
from
the discontinuity of the stress tensor.  The energy is computed
similarly in Section 4, and the expected relation between stress and
energy is found.  Fresnel drag, in the static approximation, is
considered
in Section 5, and electrostriction in Section 6.
 Estimates in Section 7 show that the Casimir energy so constructed,
even with physically required subtractions, and including both
interior
and exterior contributions, is divergent, but that if one
supplies a plausible contact term, a finite result (at least in
leading approximation) follows. This finite result agrees with that
found using $\zeta$-function regularization. (Physically, we expect
that the divergence is regulated by including dispersion.)
Numerical estimates of both the divergent and finite terms are
given in the conclusion, and comparison is made with the calculations
of Schwinger and others.  A simple estimate is given which suggests
that any macroscopic electromagnetic phenomenon such as the Casimir
effect cannot possibly supply the energy required for
sonoluminescence.
There, and in the Appendix, where we discuss the form of the
force on the surface due to the fluctuating electric and magnetic
fields, a comparison with \cite{eberlein} is made.

\section{Green's Dyadic Formulation}
We follow closely the formulation given in \cite{mds,kim}.  We start
with Maxwell's equations in rationalized units, with a polarization
source $\bf P$:  (in the following we set $c=\hbar=1$)
\begin{eqnarray}
\bbox{\nabla}\times{\bf H}={\partial\over\partial t}{\bf D}
+{\partial\over\partial t}{\bf P }&,&\qquad \bbox{\nabla}\cdot{\bf
D}=-\bbox{\nabla}\cdot
{\bf P},\nonumber\\
-\bbox{\nabla}\times{\bf E}={\partial\over\partial t}{\bf B}
&,&\qquad
\bbox{\nabla}\cdot{\bf B}=0,
\end{eqnarray}
where, for an homogeneous, isotropic, nondispersive medium
\begin{equation}
{\bf D=\epsilon E,\qquad B=\mu H.}
\end{equation}
We define a Green's dyadic $\bbox{\Gamma}$ by
\begin{equation}
{\bf E(r},t)=\int (d{\bf r}')\,dt'\, \bbox{\Gamma}({\bf r},t;{\bf
r}',t')
\cdot{\bf P(r}',t')
\end{equation}
and introduce a Fourier transform in time
\begin{equation}
\bbox{\Gamma}({\bf r},t;{\bf r}',t')=\int_{-\infty}^\infty
{d\omega\over2\pi}e^{-i\omega(t-t')}\bbox{\Gamma}({\bf r},{\bf
r}';\omega),
\end{equation}
where in the following the $\omega$ argument will be suppressed.
Maxwell's equations then become (which define $\bbox{\Phi}$)
\begin{eqnarray}
\bbox{\nabla}\times\bbox{\Gamma} =i\omega\bbox{\Phi}
&,&\qquad\bbox{\nabla}\cdot\bbox{\Phi}=0,\nonumber\\
{1\over\mu}\bbox{\nabla}\times\bbox{\Phi}
=-i\omega\epsilon\bbox{\Gamma}' &,&\qquad \bbox{\nabla}
\cdot\bbox{\Gamma}'=0,
\label{maxgreen}
\end{eqnarray}
in which $\bbox{\Gamma}'=\bbox{\Gamma}+{\bf 1}/\epsilon$, where $\bf
1$ includes a
spatial
delta function.  The two solenoidal Green's dyadics given here
satisfy the following second-order equations:
\begin{mathletters}
\begin{eqnarray}
(\nabla^2+\omega^2\epsilon\mu)\bbox{\Gamma}'
&=&-{1\over\epsilon}\bbox{\nabla}\times
(\bbox{\nabla}\times{\bf 1}),\\
(\nabla^2+\omega^2\epsilon\mu)\bbox{\Phi}
&=&i\omega\mu\bbox{\nabla}\times
{\bf 1}.
\end{eqnarray}
\end{mathletters}
They can be expanded in vector spherical harmonics
\cite{jackson,stratton}
defined by
\begin{equation}
{\bf X}_{lm}={1\over\sqrt{l(l+1)}}{\bf L}Y_{lm},
\end{equation}
as follows:
\begin{mathletters}
\begin{eqnarray}
\bbox{\Gamma}'({\bf r,r'})&=&\sum_{lm}\left(f_l(r,{\bf r}'){\bf
X}_{lm}(\Omega)
+{i\over \omega
\epsilon\mu}\bbox{\nabla}\times g_l(r,{\bf r}'){\bf
X}_{lm}(\Omega)\right),\\
\bbox{\Phi}({\bf r,r'})&=&\sum_{lm}\left(\tilde g_l(r,{\bf r}'){\bf
X}_{lm}(\Omega)
-{i\over \omega}
\bbox{\nabla}\times \tilde f_l(r,{\bf r}'){\bf
X}_{lm}(\Omega)\right).
\end{eqnarray}
\end{mathletters}
When these are substituted in Maxwell's equations (\ref{maxgreen})
we obtain, first,
\begin{equation}
g_l=\tilde g_l,\qquad f_l=\tilde f_l+{1\over\epsilon}{1\over r^2}
\delta(r-r'){\bf X}^*_{lm}(\Omega'),
\end{equation}
 and then the second-order equations
\begin{mathletters}
\begin{eqnarray}
(D_l+\omega^2\mu\epsilon)g_l(r,{\bf r}')&=&i\omega\mu\int d\Omega''\,
{\bf X}^*_{lm}(\Omega'')\cdot\bbox{\nabla}''\times{\bf 1},\\
(D_l+\omega^2\mu\epsilon)f_l(r,{\bf r}')&=&-{1\over\epsilon}\int
d\Omega''\,
{\bf X}^*_{lm}(\Omega'')\cdot\bbox{\nabla}''
\times(\bbox{\nabla}''\times{\bf 1})
\nonumber\\
&=&{1\over\epsilon}D_l{1\over r^2}\delta(r-r'){\bf
X}^*_{lm}(\Omega'),
\end{eqnarray}
\end{mathletters}
where
\begin{equation}
D_l={\partial^2\over\partial r^2}+{2\over r}{\partial\over\partial r}
-{l(l+1)\over r^2}.
\end{equation}

These equations can be solved in terms of Green's functions
satisfying
\begin{equation}
(D_l+\omega^2\epsilon\mu){ F}_l(r,r')=-{1\over r^2}\delta(r-r'),
\label{gfneqn}
\end{equation}
which have the form
\begin{equation}
{F}_l(r,r')=\left\{
\begin{array}{ll}
ik'j_l(k'r_<)[h_l(k'r_>)-Aj_l(k'r_>)],&r,r'<a,\\
ikh_l(kr_>)[j_l(kr_<)-Bh_l(kr_<)],&r,r'>a,
\end{array}\right.
\label{form}
\end{equation}
where
\begin{equation}
k=|\omega|\sqrt{\mu\epsilon},\qquad k'=|\omega|\sqrt{\mu'\epsilon'},
\end{equation}
and $h_l=h_l^{(1)}$ is the spherical Hankel function of the first
kind.
Specifically, we have
\begin{mathletters}
\begin{eqnarray}
\tilde f_l(r,{\bf r'})&=&\omega^2\mu F_l(r,r'){\bf
X}_{lm}^*(\Omega'),\\
g_l(r,{\bf r'})&=&-i\omega\mu\bbox{\nabla}'\times G_l(r,r'){\bf
X}_{lm}^*(\Omega'),
\end{eqnarray}
\end{mathletters}
where $F_l$ and $G_l$ are Green's functions of the form (\ref{form})
with the constants $A$ and $B$ determined by the boundary conditions
given below.  Given $F_l$, $G_l$, the fundamental Green's dyadic
is given by
\begin{eqnarray}
\bbox{\Gamma}'({\bf r,r}')&=&\sum_{lm}\bigg\{\omega^2\mu
F_l(r,r'){\bf
X}_{lm}(\Omega)
{\bf X}_{lm}^*(\Omega')\nonumber\\
&&\quad-{1\over\epsilon}\bbox{\nabla}\times G_l(r,r'){\bf
X}_{lm}(\Omega)
{\bf
X}_{lm}^*(\Omega')\times{\stackrel{\leftarrow}{\bbox{\nabla}}}{}'
\nonumber\\
&&\quad+{1\over\epsilon}{1\over r^2}\delta(r-r'){\bf
X}_{lm}(\Omega)
{\bf X}_{lm}^*(\Omega')\bigg\}.
\label{gam}
\end{eqnarray}

Now we consider a sphere of radius $a$ centered at the origin,
with properties $\epsilon'$, $\mu'$ in the interior and $\epsilon$,
$\mu$ outside.
Because of the boundary conditions that
\begin{equation}
{\bf E}_\perp, \quad\epsilon E_r,\quad B_r, \quad {1\over\mu}{\bf
B}_\perp
\end{equation}
be continuous at $r=a$, we find for the constants $A$ and $B$ in the
two Green's functions in (\ref{gam})
\begin{mathletters}
\begin{eqnarray}
A_F&=&{\sqrt{\epsilon\mu'}\tilde e_l(x')\tilde e_l'(x)-
\sqrt{\epsilon'\mu}\tilde e_l(x)\tilde e_l'(x')\over\Delta_l},\\
B_F&=&{\sqrt{\epsilon\mu'}\tilde s_l(x')\tilde s_l'(x)-
\sqrt{\epsilon'\mu}\tilde s_l(x)\tilde s_l'(x')\over\Delta_l},\\
A_G&=&{\sqrt{\epsilon'\mu}\tilde e_l(x')\tilde e_l'(x)-
\sqrt{\epsilon\mu'}\tilde e_l(x)\tilde
e_l'(x')\over\tilde\Delta_l},\\
B_G&=&{\sqrt{\epsilon'\mu}\tilde s_l(x')\tilde s_l'(x)-
\sqrt{\epsilon\mu'}\tilde s_l(x)\tilde s_l'(x')\over\tilde\Delta_l}.
\end{eqnarray}
\end{mathletters}
Here we have introduced $x=ka$, $x'=k'a$, the Riccati-Bessel
functions
\begin{equation}
\tilde e_l(x)=x h_l(x),\qquad \tilde s_l(x)=xj_l(x),
\end{equation}
and the denominators
\begin{eqnarray}
\Delta_l=\sqrt{\epsilon\mu'}\tilde s_l(x')\tilde e_l'(x)
-\sqrt{\epsilon'\mu}\tilde s_l'(x')\tilde e_l(x),\nonumber\\
\tilde\Delta_l=\sqrt{\epsilon'\mu}\tilde s_l(x')\tilde e_l'(x)
-\sqrt{\epsilon\mu'}\tilde s_l'(x')\tilde e_l(x),
\end{eqnarray}
and have denoted differentiation with respect to the argument by a
prime.

\section{Stress on the sphere}
We can calculate the stress (force per unit area) on the sphere by
computing
the discontinuity of the radial-radial component of the stress
tensor (see the Appendix):
\begin{equation}
{\cal F}=T_{rr}(a-)-T_{rr}(a+),
\end{equation}
where
\begin{equation}
T_{rr}={1\over2}\langle[\epsilon(E^2_\perp-E_r^2)
+\mu(H_\perp^2-H_r^2)]\rangle.
\end{equation}
The   vacuum expectation values of the product of field
strengths are given directly by the Green's dyadics computed in
Section 2:
\begin{mathletters}
\begin{eqnarray}
i\langle{\bf E(r) E(r')}\rangle&=&\bbox{\Gamma}({\bf r,r'}),\\
i\langle{\bf B(r)
B(r')}\rangle&=&-{1\over\omega^2}\bbox{\nabla}\times
\bbox{\Gamma}({\bf
r,r'})\times{\stackrel{\leftarrow}{\bbox{\nabla}}}{}',
\end{eqnarray}
\end{mathletters}
where here and in the following
we ignore $\delta$ functions because we are interested in the
{\it limit\/} as $\bf r'\to r$.
It is then rather immediate to find for the stress on the sphere
(the {\it limit\/} $t'\to t$ is assumed)
\begin{mathletters}
\begin{eqnarray}
\cal{F}&=&{1\over2ia^2}\int_{-\infty}^\infty
{d\omega\over2\pi}e^{-i\omega
(t-t')}\sum_{l=1}^\infty{2l+1\over4\pi}\nonumber\\
&&\quad\times\Bigg\{(\epsilon'-\epsilon)\left[{k^2\over\epsilon}a^2
F_l(a+,a+)+\left({l(l+1)\over\epsilon'}
+{1\over\epsilon}{\partial\over
\partial r}r{\partial\over\partial
r'}r'\right)G_l(r,r')\bigg|_{r=r'=a+}
\right]\nonumber\\
&&\mbox{}+(\mu'-\mu)\left[{k^2\over\mu}a^2
G_l(a+,a+)+\left({l(l+1)\over\mu'}+{1\over\mu}{\partial\over
\partial r}r{\partial\over\partial
r'}r'\right)F_l(r,r')\bigg|_{r=r'=a+}
\right]\Bigg\}\\
&=&{i\over2a^4}\int_{-\infty}^\infty{dy\over2\pi}e^{-iy\delta}
\sum_{l=1}^\infty{2l+1\over4\pi}x{d\over
dx}\ln\Delta_l\tilde\Delta_l,
\label{unsubstress}
\end{eqnarray}
\end{mathletters}
where $y=\omega a$, $\delta=(t-t')/a$, and
\begin{equation}
\ln\Delta_l\tilde\Delta_l=\ln\left[(\tilde s_l(x')\tilde e_l'(x)-
\tilde s_l'(x')\tilde e_l(x))^2-\xi^2(\tilde s_l(x')\tilde e_l'(x)+
\tilde s_l'(x')\tilde e_l(x))^2\right]+\mbox{constant}.
\end{equation}
Here the parameter $\xi$ is
\begin{equation}
\xi={\sqrt{{\epsilon'\over\epsilon}{\mu\over\mu'}}-1\over
\sqrt{{\epsilon'\over\epsilon}{\mu\over\mu'}}+1}.
\end{equation}

This is not yet the answer.  We must remove the term which would be
present if either medium filled all space (the same was done in the
case of parallel dielectrics \cite{sdm}).  The corresponding
Green's function is
\begin{equation}
F_l^{(0)}=\left\{\begin{array}{ll}
ik'j_l(k'r_<)h_l(k'r_>),&r,r'<a\\
ikj_l(kr_<)h_l(kr_>),&r,r'>a\end{array}\right.
\end{equation}
The resulting stress is
\begin{eqnarray}
{\cal F}^{(0)}&=&{1\over a^3}\int_{-\infty}^\infty{d\omega\over 2\pi}
 e^{-i\omega\tau}
\sum_{l=1}^\infty{2l+1\over4\pi}
\bigg\{x'[\tilde s_l'(x')\tilde e_l'(x')-\tilde e_l(x')
\tilde s_l''(x')]\nonumber\\
&&\qquad\qquad\mbox{}-x[\tilde s_l'(x)
\tilde e_l'(x)-\tilde e_l(x)
\tilde s_l''(x)]\bigg\}.
\label{vacstress}
\end{eqnarray}
The final formula for the stress is obtained by subtracting
(\ref{vacstress})
from (\ref{unsubstress}):
\begin{eqnarray}
{\cal
F}&=&-{1\over2a^4}\int_{-\infty}^\infty{dy\over2\pi}e^{iy\delta}
\sum_{l=1}^\infty{2l+1\over4\pi}
\Bigg\{x{d\over dx}\ln\Delta_l\tilde\Delta_l\nonumber\\
&&+2x'[ s_l'(x') e_l'(x')- e_l(x')
 s_l''(x')]
-2x[ s_l'(x) e_l'(x)- e_l(x)
 s_l''(x)]\Bigg\},
\label{stress}
\end{eqnarray}
where we have now performed a Euclidean rotation,
\begin{eqnarray}
y&\to& iy,\quad x\to ix,\quad\tau=t-t'\to i(x_4-x_4')
\quad [\delta=(x_4-x_4')/a],\nonumber\\
\tilde s_l(x)&\to& s_l(x)=\sqrt{\pi x\over 2}I_{l+1/2}(x),\quad
\tilde e_l(x)\to e_l(x)={2\over\pi}\sqrt{{\pi x\over 2}}K_{l+1/2}(x).
\end{eqnarray}

\section{Total energy}

In a similar way we can directly calculate the Casimir energy of the
configuration, starting from the energy density
\begin{equation}
U={\epsilon E^2+\mu H^2\over2}.
\end{equation}
In terms of the Green's dyadic, the total energy is
\begin{mathletters}
\begin{eqnarray}
E&=&\int(d{\bf r})\,U\nonumber\\
&=&{1\over2i}\int r^2
dr\,d\Omega\left[\epsilon\mbox{Tr}\,\bbox{\Gamma}({\bf r,r})-
{1\over\omega^2\mu}\mbox{Tr}\,\bbox{\nabla}\times\bbox{\Gamma}({\bf
r,r})\times\stackrel{\leftarrow}
{\bbox{\nabla}}\right]\\
&=&{1\over2i}\int_{-\infty}^\infty{d\omega\over2\pi}
e^{-i\omega(t-t')}
\sum_{l=1}^\infty(2l+1)\int_0^\infty r^2\,dr\nonumber\\
&&\times\left\{2k^2[F_l(r,r)+G_l(r,r)]
+{1\over r^2}{\partial\over\partial r}r{\partial\over\partial r'}r'
[F_l+G_l](r,r')\big|_{r'=r}\right\},
\label{fullen}
\end{eqnarray}
\end{mathletters}
where there is no explicit appearance of $\epsilon$ or $\mu$.
(However, the value of $k$ depends on which medium we are in.)
As in \cite{mds} we can easily show that the total derivative term
integrates
to zero. We are left with
\begin{equation}
E={1\over2i}\int_{-\infty}^\infty{d\omega\over2\pi}e^{-i\omega\tau}
\sum_{l=1}^\infty(2l+1)\int_0^\infty r^2\,dr\,
2k^2[F_l(r,r)+G_l(r,r)].
\label{en}
\end{equation}
However, again we should subtract off that contribution which the
formalism
would give if either medium filled all space.  That means we should
replace $F_l$ and $G_l$ by
\begin{equation}
\tilde F_l, \tilde G_l=\left\{\begin{array}{ll}
-ik'A_{F,G}j_l(k'r)j_l(k'r'),&r,r'<a\\
-ikB_{F,G}h_l(kr)h_l(kr'),&r,r'>a
\end{array}\right.
\label{rep}
\end{equation}
so then (\ref{en}) says
\begin{eqnarray}
E&=&-\sum_{l=1}^\infty(2l+1)\int_{-\infty}^\infty{d\omega\over
2\pi}e^{-i\omega\tau}
\bigg\{\int_0^a r^2dr\,k^{\prime 3}(A_F+A_G)j_l^2(k'r)\nonumber\\
&&\qquad\qquad\mbox{}+
\int_a^\infty r^2dr\,k^3(B_F+B_G)h_l^2(kr)\bigg\}.
\end{eqnarray}
The radial integrals may be done by using the following indefinite
integral for any spherical Bessel function $j_l$:
\begin{equation}
\int dx\,x^2j_l^2(x)={x\over2}[((xj_l)')^2-j_l(xj_l)'-xj_l(xj_l)''].
\end{equation}
But we must remember to add the contribution of the total derivative
term
in (\ref{fullen}) which no longer vanishes when the replacement
(\ref{rep})
is made.  The result is precisely that expected from the stress
(\ref{stress}),
\begin{equation}
E=4\pi a^3{\cal F},\qquad {\cal F}={1\over4\pi a^2}
\left(-{\partial\over\partial a}
\right)E,
\label{forderen}
\end{equation}
where the derivative is the naive one, that is,  the cutoff
$\delta$ has no effect on the derivative.

It is useful here to make contact with the formalism introduced by
Schwinger \cite{rederiv}.  In terms of an imaginary frequency $\zeta$
and a parameter $w$, he derives the following simple formula for the
energy from the proper-time formalism
\begin{equation}
E=-{1\over2\pi}\int_0^\infty d\zeta\int_0^\infty dw\, \mbox{Tr}_s G,
\end{equation}
where the trace refers to space, the Green's function is
\begin{equation}
G={1\over w+H}
\end{equation}
 and the Hamiltonian appropriate to the
two modes is (for a nonmagnetic material)
\begin{equation}
H=\left\{\begin{array}{cc}
\mbox{TE}:& \partial_0\epsilon\partial_0-\nabla^2\\
\mbox{TM}:&\partial_0^2-\bbox{\nabla}\cdot(1/\epsilon)\bbox{\nabla}
\end{array}\right.
\end{equation}

Consider the TE part (the TM part is similar, but not explicitly
considered
by Schwinger).  In terms of Green's function satisfying
(\ref{gfneqn}),
we have
\begin{equation}
E={1\over2\pi}\int_0^\infty d\zeta\int_0^\infty d w\sum_{l=1}^\infty
(2l+1)\int_0^\infty dr\, r^2 F_l(r,r;\zeta^2\epsilon+w),
\end{equation}
where the third argument of the Green's function reflects the
substitution in (\ref{gfneqn}) of
$\omega^2\epsilon\to-\zeta^2\epsilon
-w$.  We now introduce polar coordinates by writing
\begin{equation}
\zeta^2\epsilon+w=\rho^2,\quad d\zeta\,d w={1\over\sqrt{\epsilon}}
2\rho^2\cos\theta\, d\rho\,d\theta,
\end{equation}
and integrate over $\theta$ from 0 to $2\pi$.  The result coincides
with the first term in (\ref{en}).

\section{Fresnel drag}
As may easily inferred from Pauli's book \cite{pauli},
the nonrelativistic effect of material
motion of the dielectric, $\bbox{\beta}({\bf r})$, is given by the
so-called
Fresnel drag term,
\begin{equation}
E'=\int (d{\bf r}){\epsilon\mu-1\over\epsilon}\bbox{\beta}\cdot({\bf
D
\times H})=\int (d{\bf r})(\epsilon\mu-1)\bbox{\beta}\cdot({\bf
E\times
H}).
\end{equation}
To preserve spherical symmetry (of course, this is likely not to be
realistic) we consider purely radial velocities,
\begin{equation}
\bbox{\beta}=\beta\hat{\bf r}.
\end{equation}
Then, what we seek is the asymmetrical structure
\begin{eqnarray}
\hat{\bf r}\cdot\langle {\bf E(r)\times H(r')}\rangle
&=&-\hat{\bf r}\cdot\langle {\bf H(r')\times E(r)}\rangle
=-{1\over i\mu}\epsilon_{ijk}\hat{ r}_i\cdot
\Phi_{jk}({\bf r',r})\nonumber\\
&=&\omega \hat{\bf r}\cdot\sum_{lm}\bigg\{{\bf X}_{lm}(\Omega')
\times[\bbox{\nabla}\times G_l(r',r){\bf
X}_{lm}^*(\Omega)]\nonumber\\
&&\mbox{}+[\bbox{\nabla}'\times F_l(r',r){\bf X}_{lm}(\Omega')]
\times{\bf X}^*_{lm}(\Omega)\bigg\}.
\end{eqnarray}
This is easily seen to reduce to
\begin{eqnarray}
\hat{\bf r}\cdot\langle {\bf E\times H}\rangle&=&
\omega {1\over r}{\partial\over\partial r}r\sum_{lm}G_l(r',r)
{\bf X}_{lm}(\Omega')\cdot{\bf X}_{lm}^*(\Omega)\nonumber\\
&&\mbox{}-\omega {1\over r'}{\partial\over\partial r'}r'
\sum_{lm}F_l(r',r)
{\bf X}_{lm}(\Omega')\cdot{\bf X}^*_{lm}(\Omega),
\end{eqnarray}
so when $\Omega$ and $\Omega'$ are identified, and the angular
integral is carried out,
we obtain the corresponding energy for a {\it slow, adiabatic,
radially symmetric motion},
\begin{equation}
E'=\beta\int_0^\infty r^2dr\,(\epsilon\mu-1)\int_{-\infty}^\infty
{d\omega\over2\pi} \omega \, e^{-i\omega\tau}\sum_{l=1}^\infty(2l+1)
\left\{{1\over
r}{\partial
\over\partial r}r G_l(r',r)-{1\over r'}{\partial\over\partial r'}
r'F_l(r',r)\right\}\!\!\Bigg|_{r'=r}.
\end{equation}
It is clear, immediately, that if the cutoff $\tau$ is set equal to
zero, this vanishes because the integrand is odd in $\omega$; compare
to (\ref{en}).  Since the sign of $\tau$ is certainly irrelevant, we
therefore claim that in this quasistatic
approximation
Fresnel drag is absent.

Related is the Abraham value of the field momentum
\cite{breves,breves2},
\begin{equation}
{\bf G}={\bf E\times H},
\end{equation}
which gives then an extra contribution to the force density,
\begin{equation}
{\bf f}'=(\epsilon-1){\partial\over\partial t}({\bf E\times H}).
\end{equation}
However, as Brevik noted \cite{breves2}, this is also zero,
because, in the Fourier transform, successive action of the time
derivative brings down $\omega$ and $-\omega$.  So the continuing
controversy about which field momentum to use is without consequence
here.  This should be already obvious, because the energy is
well-defined,
and we have already seen that the force is related to the energy by
(\ref{forderen}).  See also the Appendix.

\section {Electrostriction}
When a dielectric medium is deformed, there is an additional
contribution
to the force density, that of electrostriction
\cite{stratton,breves},
\begin{equation}
{\bf f}_{\text{ES}}={1\over2}\bbox{\nabla}
\left(E^2\rho{\partial\epsilon\over\partial\rho}\right),
\label{fes}
\end{equation}
where $\rho$ is the density of the medium.  This term is without
effect
for computation of the {\it force\/} on the dielectric, because it is
a total derivative, yet here, where we are calculating the {\it
stress\/}
on the surface, it can be significant.  The simplest model for
describing
the density dependence of the dielectric constant is that given by
the Clausius-Mossotti equation,
\begin{equation}
{\epsilon-1\over\epsilon+2}=K\rho,
\end{equation}
where $K$ is a constant.  Consequently the logarithmic derivative
appearing in (\ref{fes}) is
\begin{equation}
\rho{\partial\epsilon\over\partial\rho}
={1\over3}(\epsilon-1)(\epsilon+2).
\end{equation}

The calculation of the electrostrictive Casimir effect for a
dielectric ball is given by Brevik \cite{breves2}.  We have
confirmed his result, and generalized it to the situation at
hand. Again, the contribution if either medium fills all space
has been subtracted. The result for the integrated stress
on the spherical cavity, after the Euclidean transformation is
performed, is
\begin{eqnarray}
F_{\text{ES}}&=&-{1\over12a^2}
\sum_{l=1}^\infty(2l+1)\int_{-\infty}^\infty
dy \, e^{iy\delta}\nonumber\\
&&\cdot\Bigg\{{(\epsilon'-1)(\epsilon'+2)\over\epsilon'}\bigg[
{A_G\over2x'}\left(x^{\prime 2}I_{l+1/2}^2(x')\right)'\nonumber\\
&&\mbox{}-x'(A_F+A_G)\int_0^{x'}d\xi\,
I_{l+1/2}^2(\xi)+x'A_G\int_0^{x'}
{d\xi\over\xi}I_{l+1/2}^2(\xi)\bigg]\nonumber\\
&&\mbox{}+{(\epsilon-1)(\epsilon+2)\over\epsilon}
\left(2\over\pi\right)^2
\bigg[-{B_G\over2x}\left(x^{ 2}K_{l+1/2}^2(x)\right)'\nonumber\\
&&\mbox{}-x(B_F+B_G)\int_x^{\infty}
d\xi\,K_{l+1/2}^2(\xi)+xB_G\int_x^{\infty}
{d\xi\over\xi}K_{l+1/2}^2(\xi)\bigg]\Bigg\}.
\label{esform}
\end{eqnarray}

\section{Asymptotic analysis and numerical results}

The result for the stress (\ref{stress}) is an immediate
generalization
of that given in \cite{kim}, and therefore, the asymptotic analysis
given there can be applied nearly unchanged.  The result for the
energy
is new, and seems not to have been recognized earlier.

We first remark on the special case
$\sqrt{\epsilon\mu}=\sqrt{\epsilon'\mu'}$.
Then $x=x'$ and the energy reduces to
\begin{equation}
E=-{1\over4\pi a}\int_{-\infty}^\infty dy\,
e^{iy\delta}\sum_{l=1}^\infty
(2l+1)x{d\over dx}\ln[1-\xi^2((s_le_l)')^2],
\label{special}
\end{equation}
where
\begin{equation}
\xi={\mu-\mu'\over\mu+\mu'}.
\label{emu}
\end{equation}
If $\xi=1$ we recover the case of a perfectly conducting spherical
shell, treated in \cite{mds}, for which $E$ is finite.  In fact
(\ref{special})
is finite for all $\xi$, and if we use the leading uniform asymptotic
approximation for the Bessel functions we obtain
\begin{equation}
E\sim {3\over64a}\xi^2.
\end{equation}
Further analysis of this special case is given by Brevik and
Kolbenstvedt
\cite{brevik}.

In general, using the uniform asymptotic behavior, with $x=\nu z$,
$\nu=l+1/2$, and, for simplicity looking at the large $z$ behavior,
we have
\begin{equation}
E\sim-{1\over 2\pi
a}{1\over\sqrt{\epsilon\mu}}\sum_{l}\nu^2\int_{-\infty}
^\infty dz\,e^{iz\nu\delta/\sqrt{\epsilon\mu}}
z{d\over
dz}\ln\left[1+{1\over16z^4}\left({\epsilon\mu\over\epsilon'\mu'}
-1\right)^2(1-\xi^2)\right],
\end{equation}
which exhibits a cubic divergence as $\delta\to0$.  To be more
explicit,
let us content ourselves with with the case when $\epsilon-1$,
$\epsilon'-1$
are both small and $\mu=\mu'=1$.  Then, the leading $\nu$ term is
\begin{eqnarray}
E&\sim&-{(\epsilon'-\epsilon)^2\over16\pi
a}\sum_{l=1}^\infty\nu^2{1\over2}
\int_{-\infty}^\infty dz\,e^{i\nu z\delta}z{d\over
dz}{1\over(1+z^2)^2}
\nonumber\\
&=&-{(\epsilon'-\epsilon)^2\over64
a}\left({16\over\delta^3}+{1\over4}
\right)\to-{(\epsilon'-\epsilon)^2\over256 a}.
\label{smalle}
\end{eqnarray}
Here, the last arguable step is made plausible by noting that since
$\delta=\tau/a$ the divergent term represents a contribution to the
surface tension on the bubble, which should be cancelled by a
suitably
chosen counter term (contact term). This argument is given somewhat
more weight by the discussion in \cite{toward}.
In essence, justification is provided there for the use of
zeta-function regularization, which directly gives the finite part
here:
\begin{equation}
E\sim{(\epsilon'-\epsilon)^2\over32\pi a}\sum_{l=1}^\infty\nu^2
{\pi\over2}={(\epsilon'-\epsilon)^2\over64 a}\left(-{1\over4}\right),
\end{equation}
because $\sum_{l=0}^\infty\nu^s=(2^{-s}-1)\zeta(-s)$ vanishes at
$s=2$.

Alternatively, one could argue that
dispersion should be included \cite{brevein,candelas,bss},
crudely modelled by
\begin{equation}
\epsilon(\omega)-1={\epsilon_0-1\over1-{\omega^2/\omega_0^2}}.
\end{equation}
If this rendered the expression for the stress finite [we consider
the stress, not the energy, for it is not necessary to consider the
dispersive factor $d(\omega\epsilon(\omega))/d\omega$ there], we
could
drop the cutoff $\delta$ and the sign of the force would be positive:
(at last, we set $\epsilon'=1$)
\begin{equation}
{\cal F}\sim+{(\epsilon_0-1)^2\over128\pi^2a^4}
\sum_{l=1}^\infty\nu^2\int_{-\infty}^\infty
dz\,{1\over(1+z^2)^2}{1\over(1+z^2/z_0^2)^2},
\label{dispforce}
\end{equation}
where $z_0=\omega_0a/\nu$.  As $\nu\to\infty$, $z_0\to0$, and
the integral here approaches $\pi z_0/2$, and so
\begin{equation}
{\cal F}\sim{(\epsilon_0-1)^2\over256\pi
a^3}\omega_0\sum_{l=1}^{\nu_c}
\nu\sim{(\epsilon_0-1)^2\over512\pi a}\omega_0^3,
\label{dispforce2}
\end{equation}
if we take as the cutoff\footnote{Inconsistently, for then
$z_0\sim1$.  If $z_0=1$ in (\ref{dispforce}), however, the same
angular momentum cutoff gives $5/12$ of the value in
(\ref{dispforce2}).}
of the angular momentum sum $\nu_c\sim\omega_0a$.
The corresponding energy is obtained by integrating $-4\pi a^2{\cal
F}$,
\begin{equation}
E\sim-{(\epsilon_0-1)^2\over256}\omega_0^3a^2,
\label{incldisp}
\end{equation}
which is of the form of (\ref{smalle}) with $1/\delta\to\omega_0a/4$.

It is rather more difficult to extract numerical results from the
formula for electrostriction, (\ref{esform}).  Indeed, Brevik
\cite{breves} only considers two special cases, $\epsilon\gg1$,
appropriate to a perfect conductor,
and $\epsilon-1\ll1$. In fact, in the latter case, corresponding to
(\ref{smalle}), he was able to consider only the single $l=1$ term
in the sum.  This is highly unreliable, as such a term may be
completely
unrepresentative (such as having the wrong sign \cite{mds}).
Because this electrostrictive stress presents new and
somewhat difficult to understand
divergences, we will defer its consideration to a later publication,
and only remark that it is highly likely to contribute a term
comparable
to the finite Casimir estimate.

\section{Conclusions}

So finally, what can we say about sonoluminescence?  To calibrate our
remarks, let us recall (a simplified version of) the argument of
Schwinger
\cite{js}.  On the basis of a provocative but
incomplete analysis he argued that
a bubble ($\epsilon'=1$) in water ($\epsilon\approx (4/3)^2$)
possessed a positive Casimir energy\footnote{Note, for small
$\epsilon-1$, Schwinger's result goes like $(\epsilon-1)$, rather
than $(\epsilon-1)^2$,
indicating that he had not removed the ``vacuum'' contribution
corresponding to (\ref{vacstress}).  This is the essential physical
reason for the discrepancy between his results and ours.}
\begin{equation}
E_c\sim{4\pi a^3\over3}\int{(d{\bf k})\over(2\pi)^3}{1\over2}k\left(
1-{1\over\sqrt{\epsilon}}\right)\sim{a^3
K^4\over12\pi}\left(1-{1\over
\sqrt{\epsilon}}\right),
\end{equation}
where $K$ is a wavenumber cutoff.  Putting in his estimate,
$a\sim 4\times 10^{-3}$cm, $K\sim2\times 10^{5}$cm$^{-1}$ (in the
UV),
we find a large Casimir energy, $E_c\sim 13$ MeV, and something like
3 million photons would be liberated if the bubble collapsed.

What does our full (albeit static) calculation say?  If we believe
the subtracted result, the last form in (\ref{smalle}), and say that
the bubble collapses from an initial radius $a_i=4\times 10^{-3}$cm
to a final radius $a_f=4\times10^{-4}$cm, as suggested by experiments
\cite{sono2}, we find that the change in
the Casimir energy is $\Delta E\sim+10^{-4}$eV.  This is far to small
to account for the observed emission.

On the other hand, perhaps we should retain the divergent result, and
put in reasonable cutoffs.  If we do so, we have
\begin{equation}
E=-{(\epsilon-1)^2\over 4}a^2K^3\sim-4\times10^5\mbox{eV},
\label{cutoff}
\end{equation}
perhaps of acceptable magnitude, but of the {\it wrong\/} sign.
(The emission occurs {\it during\/} the collapse.)
The same conclusion follows if one uses dispersion, as
(\ref{incldisp})
shows.

So we are unable to see how the Casimir effect could possibly
supply energy relevant to the copious emission of light seen in
sonoluminescence.  Of course, dynamical effects could change this
conclusion, but elementary arguments suggest that this is impossible
unless ultrarelativistic velocities are achieved.  For example,
consider
the Larmor formula,
appropriate to dipole radiation; it gives the power radiated:
\begin{equation}
P={2\over3}{(\ddot{\bf d})^2\over c^3},
\label{larmor}
\end{equation}
where $\bf d$ is the dipole moment.  If our bubble, with $N$ atoms,
coherently emits radiation, we expect
\begin{equation}
|\ddot{\bf d}|\sim{N d_a\over\tau^2},
\end{equation}
where $d_a$ is an atomic dipole moment, and $\tau$ a characteristic
collapse time for the bubble.  Thus the energy emitted during one
collapse of a bubble in water is
\begin{equation}
E\sim4\pi\alpha\hbar c
\left(10^{23}\left({a\over\text{cm}}\right)^3\right)^2
{(d_a/e)^2\over(c\tau)^3}.
\label{estimate}
\end{equation}
So with $a\sim10^{-3}$ cm, $\tau\sim10^{-5}$ s (suggested by
experiments
\cite{sono2}), and $d_a\sim10^{-8} e$-cm,
we get an energy of only $E\sim 10^{-10}$ eV.  This is in spite of
the
assumption of coherent radiation.   Note that in (\ref{estimate}),
$\tau$
would have to be $\sim 10^{-11}$ s (which is the upper bound to the
observed flash duration) to yield a total energy of 10 MeV;
this would correspond to a velocity across the bubble of $10^8$ cm/s,
well in excess of the speed of sound, thus precluding the presumed
coherent
radiation process.

 We therefore believe that in Eberlein's calculation \cite{eberlein}
there is an implicit assumption of
superluminal velocities.\footnote{Indeed, if one follows Eberlein
and uses $\gamma\sim 1$ fs (though the experimental value seems to be
closer to 10 ps) in her model profile, one finds the maximum speed of
the
bubble surface to exceed the speed of light by almost two orders of
magnitude.  Actually, even with such a small $\gamma$, we find her
result yields an energy output of only $10^{-3}$ MeV, insufficient
to explain sonoluminescence.  Eberlein agrees with these points, and
no longer stands by this model \cite{retract}.}  We note that the
short wavelength result of Eberlein, (4.7) of the first reference in
\cite{eberlein} or (10) of the second, can be cast in the dipole form
(\ref{larmor}) by integrating by parts.  Up to factors nearly equal
to one,
\begin{equation}
\left(d\over e\right)_E\approx a{\dot a\over c},
\end{equation}
where $a(t)$ is the bubble radius.  Because $\dot a/c<1$, we find
that
emission energy of 10 MeV requires a time scale of $\tau_E\sim
10^{-17}$
s.  This seems to us an implausibly short scale unless remarkable
relativistic phenomena are involved.  [Incidentally, the magnitude
of our cutoff estimate, (\ref{cutoff}), also agrees with
(\ref{larmor})
if $K\sim 1/\tau$.  This demonstrates that there is nothing classical
about the estimate (\ref{estimate})].

The only plausible origin of such short time scales lies in the
formation
of a shock.  Such a model was sketched by Greenspan and Nadim in
\cite{theory}.  But this picture has nothing to do with quantum
radiation.

Recent experimental results have made it even more difficult to
accommodate
any explanation based on macroscopic considerations.  In particular,
Hiller and Putterman \cite{isotope} have found a remarkably strong
isotope effect when water (H$_2$O) is replaced by heavy water
(D$_2$O),
where the dielectric properties change by no more than 10\%.  This,
together with the already known strong temperature dependence, and
strong dependence on gas concentration and the gas mixture, may
rule out Casimir effect explanations entirely.
Yet the subject of vacuum energy is sufficiently subtle that
surprises could be in store.
\section*{Acknowledgments}
We thank the US Department of Energy for partial financial support of
this
research.
  We are happy to acknowledge
 useful conversations and correspondence with C. Bender, I. Brevik,
C.
Eberlein, L. Ford, and D. Sciama.
We dedicate this paper to the memory of Julian Schwinger, who taught
us
so much, and first stirred our interest in this subject and then
revived that
interest, with his remarkable suggestion that the Casimir effect is
behind
sonoluminescence.  Would that it be so!

\appendix
\section*{Discussion of form of force on surface}
There seems, in the literature, to be some confusion about the
correct
form for the stress on a surface due to electromagnetic fields
(here,
of course, we are interested in vacuum expectation values of those
fields).
The definitive discussion seems to be given in Stratton
\cite{stratton}.
We also  refer here to the manuscript in process by one of us
\cite{embook}.

In the text, we computed the force on the surface by considering the
discontinuity of the stress tensor,
\begin{equation}
T_{nn}={1\over2}\epsilon(E_\perp^2-E_n^2)
+{1\over2}\mu(H_\perp^2-H_n^2),
\end{equation}
across the surface, where $\bf n$ denotes the direction normal to the
surface,
and $\perp$ directions tangential to the surface. This follows
directly
from a consideration of the interpretation to $T_{nn}$ as the flow of
$n$th component of momentum in the direction $\bf n$.   Because of
the
boundary conditions that
\begin{equation}
{\bf E}_\perp,\quad D_n,\quad {\bf H}_\perp,\quad B_n
\end{equation}
be continuous,
the stress on the surface is
\begin{equation}
{\cal
F}=T_{nn}(-)-T_{nn}(+)={1\over2}\left[(\epsilon'-\epsilon)E_\perp^2
-\left({1\over\epsilon'}-{1\over\epsilon}\right)D_n^2
+(\mu'-\mu)H_\perp^2-\left({1\over\mu'}
-{1\over\mu}\right)B_n^2\right],
\label{ststress}
\end{equation}
in terms of fields on the surface, and where a prime denotes
quantities
on the ``$-$'' side of the surface.
This is obviously equivalent to the following
form for the force density,
\begin{equation}
{\bf
f}=-{1\over2}\left(E_\perp^2\bbox{\nabla}\epsilon-D_n^2
\bbox{\nabla}{1\over\epsilon}
+H_\perp^2\bbox{\nabla}\mu-B_n^2\bbox{\nabla}{1\over\mu}\right)
=-{1\over2}(E^2\bbox{\nabla}\epsilon
+H^2\bbox{\nabla}\mu),
\end{equation}
which is just what is obtained from the stationary principle for the
energy
\cite{embook}.

The controversy seems to center around the additional ``Abraham''
term
\begin{equation}
{\bf f}'=(\epsilon-1){\partial\over\partial t}({\bf E\times H}).
\label{abforce}
\end{equation}
(Henceforward we restrict ourselves to nonmagnetic material,
$\mu=1$.)
As noted in Section 5, this makes no contribution to Casimir effect,
because the vacuum expectation value is stationary.  Furthermore,
the existence of such a term is dependent upon the (essentially
arbitrary) split between field and particle momentum.  The Minkowski
choice for field momentum
\begin{equation}
{\bf G}_M={\bf D\times H}
\end{equation}
would not imply this additional force density.  The analysis of
experimental
data given by Brevik \cite{breves}, however, seems to favor the
Abraham
value.

If we were calculating a net {\it force\/} on the surface,
(\ref{abforce}) would indeed give a further contribution to the
force beyond that given by (\ref{ststress}).  Through use of
Maxwell's
equations, we easily find
\begin{equation}
f'_n=(\epsilon-1)\left[-{1\over2\epsilon}\nabla_n
B^2-{1\over2}\nabla_nE^2
+{1\over\epsilon}\bbox{\nabla}\cdot({\bf
B}B_n)+\bbox{\nabla}\cdot({\bf
E}E_n)\right].
\end{equation}
If $\bf n$ were a fixed direction, the volume integral of this force
density would turn into a surface integral, and the result given by
Eberlein follows,
\begin{equation}
F'_n={1\over2}\int
dS\,\left[\left({1\over\epsilon}-{1\over\epsilon'}\right)
(B_n^2-B_\perp^2)+(\epsilon-\epsilon')E_\perp^2
-\left({1\over\epsilon}-{1\over\epsilon'}\right)
\left(1-{1\over\epsilon}
-{1\over\epsilon'}\right)D_n^2\right].
\end{equation}
But this result cannot be used to compute the stress.  Thus, the
formula (C5) given in Appendix C of the first paper in
\cite{eberlein}
is wrong, and, accordingly, so is (3.18) there.\footnote{The
first derivation there is based on the formula
for the force given in the following paragraph, while the second
is based on an obviously incorrect extrapolation from the vacuum
stress tensor, which of course gives vanishing stress.}

Finally, we note there is yet another formula for the force on a
dielctric given in terms of polarization charges and currents,
\begin{equation}
{\bf F}=
\int(d{\bf r})\left[\rho_{\text{pol}}{\bf E}+{1\over c}{\bf
j}_{\text{pol}}
\times {\bf B}\right],
\label{polforce}
\end{equation}
where
\begin{equation}
\rho_{\text{pol}}=-\bbox{\nabla}\cdot{\bf P},\quad {\bf
j}_{\text{pol}}
={\partial\over\partial t}{\bf P}+c\bbox{\nabla}\times{\bf M},
\end{equation}
with the polarization and magnetization fields  given by
\begin{equation}
{\bf P}={\bf D-E}=\left(1-{1\over\epsilon}\right){\bf D},
\quad {\bf M}={\bf B-H}=(\mu-1){\bf H}.
\end{equation}
Again, it is easy to show that if one is calculating the force in
a fixed direction, so one can freely integrate by parts,  for a
nonmagnetic medium, we recover the expected force including the
Abraham term:
\begin{equation}
{\bf F}=\int(d{\bf r})\,\left[-{E^2\over2}\bbox{\nabla}\epsilon
+{1\over c}{\partial\over\partial t}(\epsilon-1){\bf E\times
B}\right].
\end{equation}
But the integrand in (\ref{polforce}) is not interpretable as a force
density from which the stress may be computed.
In effect, it is that interpretation that \cite{eberlein} uses.

\end{document}